\documentstyle[aps,preprint]{revtex}
\tightenlines
\begin{document}

\title{Using microsimulation feedback for trip adaptation for
realistic traffic in Dallas}

\author{Kai Nagel, Christopher L Barrett\\
~\\
Los Alamos National Laboratory, TSA-DO/SA MS M997, Los Alamos NM
87545, U.S.A., kai@lanl.gov, barrett@tsasa.lanl.gov\\
and\\
Santa Fe Institute, 1399 Hyde Park Rd, Santa Fe NM 87501,
kai@santafe.edu\\
~\\
\today
}

\maketitle

\begin{abstract}
This paper presents a day-to-day re-routing relaxation approach for traffic
simulations.  Starting from an initial planset for the routes, the
route-based microsimulation is executed.  The result of the microsimulation
is fed into a re-router, which re-routes a certain percentage of all trips.
This approach makes the traffic patterns in the microsimulation much more
reasonable.  Further, it is shown that the method described in this paper
can lead to strong oscillations in the solutions.
\end{abstract}

\section{Introduction}

TRANSIMS is a multi-year project funded mostly by the American Federal
Highway Administration with the purpose of developing new methods for
transportation planning.  Typical application examples are, say, the
impact of introducing a public transit system to a city, or the impact
of converting vehicle roads into pedestrian zones.  When dealing with
questions like that, there is wide agreement that simulation models
are currently the only approach available which is able to deal with
complex features of the real world.

TRANSIMS is designed in a way that it incorporates all modes of
transportation, including light rail, buses, bicycles, pedestrians,
etc.  Yet, the quantitatively most important part of transportation
certainly is individual vehicular traffic, which is the reason why the
TRANSIMS project started with it.  As a result, this paper treats
car traffic only.

For transportation planning questions like the above, the first and most
important feature is to predict delays (e.g.\ traffic jams) correctly,
i.e.\ in the right places and at the right times.  For example, when adding
a lane to a freeway, one needs to know where congestion gets better and
where it gets worse as a result.  Note that most analysis depends on
knowing this; even air pollution models will be rather wrong if the traffic
model predicts congestion in wrong places.  Also note that optimization
might be based on these results as a next step.

Two features of a traffic simulation are the most important ones for
achieving the prediction of delays: (i) realistic traffic flow
dynamics; (ii) a realistic way of ``driving'' the traffic, i.e.\ a way
of telling the vehicles or the traffic streams where to go.  This
paper deals with the second aspect; elements of the first part can be
found in, e.g.,~\cite{Nagel:Schreckenberg,Nagel:flow,Pieck:etc:flow}.

A conventional way of driving traffic simulation models are turn
counts.  Here, for each intersection and each incoming direction, a
(possibly time dependent) table contains the information how many of
the vehicles go left, straight, right, etc.  It is fairly obvious that
this approach is useless for planning questions such as the above.
The most extreme example showing this is that, after the addition of a
new road, there would be no turn counts directing traffic on it.
Besides this, there is also a data collection problem.  The cost of
collecting turn counts for all intersections in a city, possibly for
different scenarios, is prohibitive.

For that reason, TRANSIMS uses individual route plans, i.e.\ each
individual vehicle in the simulation ``knows'' the sequence of streets
it is intending to use.  Note that this makes a {\em
micro\/}simulation, i.e.\ a simulation which resolves each individual
vehicle, an absolute requirement.  A necessary input information for
this is to have origin-destination (OD) matrices available, i.e.\
(possibly time-dependent) tables telling how many trips are made from
each possible origin to each possible destination of a city.  Although
most cities have such tables, derived from more conventional methods,
most traffic practitioners will also admit that they are rather far
off the real numbers, often by 30\% or more~\cite{Cervenka:personal}.
This indicates that data collection for these tables is again a
problem.  In addition, OD matrices are also subject to change under
infrastructure changes, although to a lesser degree as turn counts.
For example, the introduction of a public transit system may leave a
car at home which will then be used for other trips.

All this means that OD matrices cannot be the proper solution for a
transportation planning model.  The TRANSIMS design for that reason
starts with demographic data.  It first derives ``synthetic''
households from this data, with activites and locations of activities.
These activities are then put together (``chained''), their
transportation is planned and finally executed in the microsimulation.

The TRANSIMS project uses example cases (called case studies) in order
to remain focussed on real world issues and problems.  The current
case study is located in the Dallas/Fort Worth area and is done in
collaboration with the responsible Municipal Planning Organization
(MPO), which is the North Central Texas Council of Governments
(NCTCOG).  Since for some of the above modules only very preliminary
versions are available, the Dallas case study focusses on the
microsimulation and some aspects of route planning.  For that purpose,
TRANSIMS actually uses the NCTCOG trip table, knowing that it is most
probably wrong.  The focus of the case study is, in consequence, the
question if, given a trip table, the (preliminary) route planning
module and the microsimulation module of TRANSIMS can generate
reasonable traffic patterns.  Yet, it should be kept in mind that the
TRANSIMS design will ultimately go beyond using OD tables as starting
point.  For more information on the TRANSIMS case study see
Ref.~\cite{Beckman:case-study}.

\section{Traditional trip assignment}

Before we start describing the TRANSIMS method of how to proceed from
a given OD matrix, let us review the traditional approach.  The
traditional method of trip assignment from an OD matrix, called
dynamic assignment, is some variation of the following
method~\cite{Sheffi,Zallmann}.  

The first part is the initial allocation:\begin{itemize}

\item[(0.)]
Calculate link travel times from free speeds and link lengths for the
empty network.  Link travel times will be used as the ``cost
function'' throughout this paper.

\item[(1.)]  Select one of the OD streams.  Optimally route a
fraction, $1/k$, of that stream based on the cost function.  The cost
(i.e.\ travel time from origin to destination) for a stream is the sum
of all link travel times for the links it uses.

\item[(2.)]  Re-calculate the cost function (i.e.\ the link travel
times; see below) for each link based on the streams so far allocated.

\item[(3.)]
Go to step~1 until all trips are routed.

\end{itemize}

This initial assignment is often followed by an adjustment process.
For this, the link costs for the 100\% full network are calculated and
then some fraction of the trips are taken off the network and
re-routed, based on that cost function.

%%  It can be shown that the general problem is mathematically well
%%  behaved (i.e.\ convergence happens under a wide range of
%%  circumstances~\cite{Moll:personal}), and selecting a method is thus a
%%  question of convenience, personal taste, and computational speed.

The cost functions (link travel times) in this method are
traditionally dependent on demand only, i.e.
\[
ttime(link) 
= {length(link) \over speed(link)}
= {length(link) \over f(demand(link))} \ ,
\]
where $ttime(link)$ is the link travel time for that specific link,
$length(link)$ is the length of that link, $speed(link)$ is the speed
on that link, and $demand(link)$ is the number of vehicles which
intend to use that link during a given time period.

$speed = f(demand)$ is a monotonously decreasing function.  It is
usually defined in terms of the ratio between flow demand (or {\em
volume demand}) and {\em capacity\/} and is then called the
V/C-ratio.  For a low V/C-ratio, speed is close to the free speed of
the link; for a high V/C-ratio, speed is set to a low value, say
1~km/h.

Although these procedures are often not time dependent, it is easily
imaginable to use time dependent OD matrices.  The single most
important point where even the time-dependent methods break down is
after the onset of congestion, i.e.\ when demand for a certain part of
the network becomes higher than capacity.  The reason is that all
traditional assignment methods assume that all demand can always be
cleared by the links.  When V/C is much larger than one, speed will be
very low, but the method still assumes that the amount V of vehicles
will leave the link during the time period, however large V is. This
is clearly inconsistent with the actual traffic
dynamics~\cite{Johnson:inconsistent}: In reality, $V/C > 1$ implies a
queue built-up (congestion), and the link travel times become history
dependent: Just after the onset of congestion, the link travel time is
still fairly low; when a $V/C>1$ condition has existed for a long
time, the link travel time includes all the waiting time in the queue
waiting to enter the link and will thus be fairly
high~\cite{bottleneck:econ}.

The technical problem here is that keeping track of congestion
built-up demands a different view of the dynamics than traditional
assignment usually has, and this turns out to be a rather difficult
problem for the traditional methods.

\section{Trip adaptation via microsimulation feedback}

It seems that currently the only way to consistently deal with these
problems are traffic microsimulations.  Here, each individual vehicle
follows the assigned route from the assignment process, and the link
travel times (cost function) now come from the microsimulation, which
automatically calculates dynamically correct queue built-up delays.
In order to just get the queue built-up right, very simplified
microsimulation models will probably be
sufficient~\cite{Simao:queueing,Gawron:simple,Nagel:Simon}; the
TRANSIMS microsimulation is much more realistic and also includes
complications such as speed limits, turn pockets, signal phasings,
more realistic intersection behavior, etc.  How far these real world
additions change the outcome is subject to research; it is certainly
imagineable that they do: for example, a car making an unprotected
left turn against heavy traffic will have a much higher delay on a
link than a car just going straight, an effect which is only captured
with realistic intersection dynamics.

This paper presents a certain method of how the microsimulation
results can be fed back into the planning (assignment) process.  This
method matches the traditional assignment process except that it
replaces the traditional way of calculating the cost function by the
microsimulation.  In other words: the microsimulation output {\em
is\/} the cost function.  See also
Refs.~\cite{Nagel:NRW,VanAerde:96:INTEGRATION}.

The procedure used in this paper is to run the microsimulation on a
given planset (the set of all plans calculated by the planning
process), then re-route a certain fraction of the trips based on the
microsimulation result, then run the microsimulation again, etc. (see
Fig.~\ref{procedure}).  Note that in this set-up, drivers cannot
change their behavior during the microsimulation, i.e.\ during
driving.  Also note that this re-planning method uses ``old''
information as the basis of the re-routing, i.e.\ the effect of other
trips being re-routed simultaneously is not considered during the
re-routing calculation.  A ``story'' of this behavior is that each
driver writes down a sequence of roads she wants to use {\em before
starting to drive} (planning phase).  All drivers then execute these
plans without the possibility to change their mind (microsimulation
phase).  Then, say over night, a certain fraction of these people has
an opportunity to change their sequence of roads (re-planning phase),
then the microsimulation is executed again according to the
pre-calculated plans, etc.  A discussion of this is offered further
down.

This relaxation procedure needs an initial planset which has to be
generated without any microsimulation information because none is
available at that point.  This initial planset for the results
presented in this paper is generated with a variation of the
traditional assignment method with the only exception that individual
trips are routed instead fractions of streams.  We expect the general
results to be independent of the initial planset.  For that reason,
generation of the initial planset becomes an algorithmical problem
(find the initial planset which is as close as possible to the proper
solution, thus decreasing the necessary number of iterations), and
this is not part of the present paper.  For more information, see
Ref.~\cite{Marathe:planner}. 

We now continue to describe the re-routing method once the initial
planset and the initial microsimulation based on this planset have
been run.  The particular re-routing (re-planning) method used for the
results in this paper is a time dependent optimal shortest path
algorithm based on the latest microsimulation result, with the
following technical details and additions:\begin{itemize}

\item
The time dependence is done with 15-minute time bins.  That is, all
link travel times between, say, 8am and 8:15am are averaged, and that
average is used for all trips planning to enter the link in that time
period. 

\item
Remember that each traveler during the re-routing procedure uses link
travel times provided by the microsimulation.  Yet, instead of using
the correct values, each traveler uses a $\pm 30\%$ individually
distorted view of the link travel times.  Technically, for each
traveler a random number between 0.7 and 1.3 is drawn for each link
and this is multiplied with the average link travel time from the
microsimulation.

The reason for this is that the method presented here has the tendency
to create oscillations in the sense that optimal routing algorithms
put all trips on routes which have only small advantages.  Distorting
the link view for each individual traveler reduces this problem.  For
further details, see below.

\item
Tests resulted in the observation that the approach, taken literally,
did not deal very well with very fast congestion built-up, i.e.\ the
algorithm routed trips via links which just became congested.  This is
in part to be expected, since the microsimulation output reports the
link travel times for vehicles which {\em left\/} the link during a
certain time interval, whereas the re-routing procedure uses that same
time information for trips {\em entering\/} during that time interval.

We used a 900~second time shift to compensate for this problem.  That
is, a trip planning to enter a link at, say, 8:01am uses the average
link travel time information between 8:15am and 8:30am as the basis
for its decision if it wants to use that particular link.  Tests with
a 450~second time shift showed that that was not enough to deal with
some particularly quickly arising congestion.

\item
An additional feature is a certain demand ``elasticity''.  If the
result of the above route planning process includes a link whose
expected speed is less than 1~meter/second, then this trip is
deleted as ``unplannable''.  The reason for introducing this is that
the original trip tables seem to have too many trips going out of
certain residential areas during the time period under consideration
--- if these trips need forever to get out, they will probably take
place at a different point in time.

\end{itemize}

\section{Description of the simulation set-up}

All results presented in this paper are based on the following
data/parameters:\begin{itemize}

\item
The road network is the so-called local streets network for the case
study.  It includes {\em all\/} streets inside an approximately
5~miles $\times$ 5~miles study area (or region of interest).  Outside
the study area, fewer and fewer roads are included with increasing
distance from the study area.  A view of the whole network can be
found in Fig.~\ref{network}.  This street network is provided by
NCTCOG.  It includes signal timings for the signalized intersections
inside the study area.

For the results presented here, both the planner and the re-planner
operate on that whole road network, whereas the microsimulation only
operates on the study area.

\item
The origin-destination relations used in this paper are modified
version of trip tables provided by NCTCOG.  These trip tables contain
about 10~million trips for a 24~hour period of the Dallas/Fort Worth
area. 

\item
All simulations here are based on all trips which start between 5am
and 10am.  The planner which generates the initial planset generates
route plans for all these activities, but retains only those routes
which go through the study area.  Those were about 300\,000 plans.
All simulations were started at 7am.  They were run until 12noon in
order to observe the discharging behavior of the road network when no
more new trips were added.

\item
The microsimulation logic is based on the cellular automata technique
of Ref.~\cite{Nagel:Schreckenberg}.  For the velocity update, a
randomization value of $p=0.2$ was chosen, yielding maximum average
flows of approximately 2000~vehicles/hour/lane, which is about
realistic.  The lane changing rules are a multilane extension of the
symmetric two-lane rules of Ref.~\cite{Rickert:etc:twolane}.
Yield signs, stops signs, left turns against oncoming traffic, etc.,
are essentially coded according to one unifying logic: ``Interfering''
lanes (i.e.\ lanes which have priority) are identified, and the
movement is only accepted if the gap on all interfering lanes is
larger than $v_{max} = 5$.  A publication on the details of the
TRANSIMS microsimulation driving logic is in
preparation~\cite{Pieck:etc:flow}; we expect the overall results of
this paper to be independent of these details.

\item
Plan-following necessitates that vehicles are in the correct lanes at
intersections.  For example, a vehicle with the intention of a right
turn should be in one of the lanes which actually allow a right turn.
This is achieved by overriding some of the general lane changing logic
by plan following necessecities.  It is clear that, for whatever
lane-changing-for-plan-following logic, one will have to accept one of
two options: Either (i)~some vehicles get ``lost'' because they do not
make it into one of the correct lanes, or (ii)~intersections may
deadlock easily because too many vehicles for a certain turn are
blocking {\em all\/} lanes, not advancing until the correct lane has
an opening.  The current TRANSIMS microsimulation chooses option (i),
i.e.\ it accepts lost vehicles.  The amount of lost vehicles is also a
measure of the ``reasonableness'' of the planset.  Again, the
technical details of this will be treated in a different publication.

\end{itemize}

\section{Results}

A view of the microsimulation at 10:00am based on the initial planset
are shown in Fig.~\ref{0-iteration}.  It is clearly visible that there
are too many vehicles in the residential areas.  These vehicles queue
up and occupy large amounts of the residential and minor streets.
Even at 12noon, long after the last plan has started, there are still
many of these jams left, i.e.\ the simulation does not discharge its
vehicles.

Further inspection reveals that this is the result of deadlocks, which are
artifacts of certain driving rules of the
simulation~\cite{Rickert:Nagel:DFW}.  The two generic situations leading to
deadlocks are shown in Fig.~\ref{deadlocks}.  Both deadlocks could be
resolved if drivers would follow their plans less ``stubbornly'', i.e.,
after having unsuccessfully waited for a certain movement for a certain
time period, they should just do something else.  The situation shown in
the right of Fig.~\ref{deadlocks} could also be resolved if vehicles could
make left turns against oncoming traffic when that traffic is not moving.
The current gap acceptance logic demands a gap larger than $v_{max}$ in all
interfering lanes in order to allow a movement across an intersection.
Both changes will be investigated in future versions of the
microsimulation.

Yet, we found it interesting to investigate the effect of re-routing
even with a potentially deadlocking microsimulation.  The question
here is in how far routing adjustments can compensate for certain
artifacts in the microsimulation (in this case the possibility of
deadlocks). 

Fig.~\ref{1-iteration} shows the result after the first iteration,
after re-planning 20\% of the trips.  It is clear that many of the
residential area jams have decreased or even completely vanished.

Fig.~\ref{10-iteration} shows the result after the 10th iteration,
where the respective re-planning fractions have beeen 20\%, 10\%,
10\%, 10\%, 10\%, 5\%, 5\%, 5\%, 5\%, 5\%.  All residential jams have
completely vanished; what is left are very busy freeways and some
queues at traffic lights.  It is clear that this result is much more
``reasonable'' than the starting solution.  Quantitative comparisons
with reality are in preparation and will be the subject of a later
publication.  For the enjoyment of the reader, the situations at 8am
and 9am are shown in Figs.~\ref{8am} and~\ref{9am}.

In the analysis of what has happened it is clear that some congestion
is reduced because trips are re-routed through less congested areas.
Yet, there are two additional important effects:\begin{itemize}

\item
``Elasticity'' as explained above deletes a certain number of trips.
The overall number of deleted trips in all 10~iterations because of
this criterion was 5993, that is about 2\% percent of all trips from
the initial planset.

\item
Remember that the microsimulation runs on a smaller region than the
route planner.  For links with no information from the
microsimulation, the re-router assumes that they are empty, i.e.\ that
they can be travelled fast.  A result of this is that long distance
trips which use the freeways through the study area get ``pushed out''
of the area, i.e.\ re-routing puts them on other freeways which avoid
the study area.  This is justified because we had indication that the
number of trips for the region of interest was too high anyway.  Thus,
some ``automatic'' mechanism to reduce the number of trips in the
study area seemed desirable.

\end{itemize}
Overall, the number of trips going through the study area starting
between 5am and 10am as a function of the iteration is shown in
Table~\ref{written}.  After the 10th~iteration, about 10\% less trips
than initially go through the study area.

Another quantitative criterion of the success of the re-routing is the
number of lost vehicles.  Remember that ``lost'' vehicles are vehicles
in the microsimulation which did not make it into the correct lane to
execute an intended turning movement and thus went into a wrong link.
Occasions of such events are counted in the microsimulation and the
vehicles are then taken out of the simulation since the current
microsimulation does not allow on-line re-routing.
Table~\ref{written} also contains the number of lost vehicles for each
iteration.  The percentage of lost vehicles decreased from about 16\%
in the initial microsimulation to less than 6\% in the 10th iteration.

\section{Oscillations}

One prominent feature of this method of re-routing are oscillations.
The generic mechanism is easy to explain: Assume there are two routes,
I~and II, with identical characteristics, both leading from A to B.
Now assume that there is more traffic on route~I, i.e.\ route~I is
slower.  A deterministic optimizing re-router of the type used in this
paper would therefore re-route all trips that it re-routes between A
and B to route~II.  The result can be that in the following
microsimulation there is now more traffic on route~II.  In
consequence, in the next iteration the planner will route more traffic
on I, etc., causing oscillations between I and II.  Note that this is
nothing unusual for a time-discrete delay method.

Figs.~\ref{oscillation4} and~\ref{oscillation5} shows an instance of
such behavior.  Shown are {\em difference\/} plots of two consecutive
iterations.  Locations marked in white mean that density {\em
de\/}creased in this location; black means that density {\em
in\/}creased here.  One notes in several locations, especially around
the large intersection in the middle, that the system displays
systematic oscillations of the type explained above.

Remember that for each re-routing of a trip we are already using a
$\pm 30$\% individually distorted view of the link travel times.
Without this, the oscillations are much stronger.  Note that in
general this distortion only reduces the amplitude of the fluctuations
but does not dampen them out. As the most extreme example to make this
point consider an example where, in a given iteration, the planner can
choose between two different routes which have, in the current
iteration, a more than 30\% travel time difference.  In spite of the
distortion, the planner will allocate all trips on the faster route.
If this allocation now leads to an inversion of the travel time
difference, i.e.\ this route now becomes more than 30\% slower, then
the oscillation will not dampen out.  Similar, but somewhat more
complicated examples can be constructed for smaller travel time
differences.

Although the phenomenological behavior of the oscillations is easy to
explain, finding a good solution is harder for the lack of a good
theory.  For the results shown in Fig.~\ref{10-iteration} to~\ref{9am}, a
heuristic approach was used: When an oscillation became strongly
visible, the latest iterated planset was discarded and replaced by
another one with a lower re-routing percentage.  This is how the above
sequence of re-routing percentages was constructed.

\section{Computational considerations}

Investigations such as the one outlined in this paper face two
computational constraints: (i)~The computational hardware should still
be affordable for the Planning Organizations who will finally use
them.  (ii)~An iteration project such as the one presented here is
absolutely necessary in order to obtain at least reasonable results.
Computations of larger geographical areas and faster turn-around times
would be highly desirable.

The current TRANSIMS microsimulation uses distributed workstations
coupled via optical LAN using PVM.  The results presented here have
been obtained from runs using 5~Sparc5 CPUs in parallel.  The
re-router as well as pre- and postprocessing routines run on single
CPUs.  The break-down of the computing times of a single iteration is
as follows:

\halign{&#\hfil\cr
\qquad Re-router: & 1 - 3 hours, depending on the re-routing fraction \cr
\qquad Pre-processor: & 2 hours \cr
\qquad Microsimulation 7am-12am: & 5.5 hours \cr
\qquad Post-processor: & 1 hour \cr
\qquad Sum: & 9.5 - 11.5 hours \cr
}

In consequence, for 10~iterations one needs at least five days on the
described hardware; more in practice because of the heuristic way
described above in order to obtain the re-routing fraction.  Also,
several relaxation series preceded the one shown in this paper.
Overall, about 25~days of continuous computing time were needed, about
half of it on a single CPU and half of it on 5~parallel CPUs.  Faster
computing techniques on faster hardware are thus under consideration.
Yet, it is unclear how much faster the current microsimulation
technique can get on our hardware.  We know that simplified
implementations can run the same geographical area with the same
computing speed on a {\em single\/} CPU~\cite{Rickert:Nagel:DFW}, and
we know that that implementation is reasonably close to the fastest
implementations known~\cite{Nagel:Schleicher}.  That means, the
expected upper limits of computational speed improvements would be a
factor of five.  Yet, the microsimulation described here is much more
realistic then those mentioned above, and it is unclear how much of
the speed loss has to be contributed to that realism and the data
structure overhead associated with it.

Also, passing information using plain ASCII files (as TRANSIMS
currently does) poses considerable strains on the disk space.  The
original origin-destination information contains 10~million trips; in
the format currently used in TRANSIMS more than 1~GByte are needed for
that file.  A route-plans file for 5am to 10am containing all plans
going through the study area (300\,000~routes) is ca.\ 250~MByte long
(80~MByte compressed).  The microsimulation output used for this study
occupies, in the current format, approximately 50~MByte.  Multiplying
all this with 10~iterations plus the base case, one ends up with
approx.\ 2.5~GByte of disk space which are necessary for this study.
Methods to compress the files are under consideration; for example,
the routing sequence of the plans file can be compressed by a factor
of~50 using intelligent compression methods~\cite{Bush:personal}.

\section{Discussion}

One could attempt to cast the method described in this paper both in
behavioral and economics terms.  The method would then correspond to a
certain percentage of people, say 1\%, changing their behavior over
night.  (The higher re-routing percentages used in the first couple of
iterations could then be justified as a computational trick: Using 1\%
from the beginning would ultimately lead to the same overall result,
but only after many more iterations.)  In that interpretation, 1\% of
all agents would check during an over-night calculation if they could
have done better by chosing a different route, and if so, that new
route will be chosen in the future.  Note that this corresponds
roughly to a Nash equilibrium definition: The iterations would relax
if eventually {\em nobody\/} could benefit from such a routing change,
i.e.\ an individual agent selected for re-routing decides that her
current route is the best she can do.  It is unclear if (and
improbable that) the method described in this paper actually achieves
such a strong convergence.  The question if such an interpretation
could be justified {\em in the average\/} will be considered in future
work.

The re-routing method described in this paper uses {\em global
average\/} information, i.e.\ for the re-planning of each trip a
global view of the past performance of the network is available, but
this view is averaged over 15~minute bins.  Note that this introduces
two artifacts compared to the real world: (i) Nobody has complete
network performance information.  Such information could though be
imagined as the result of future traveler information systems.
Modifications of the approach presented here could thus yield
information on how such systems could change the system. (ii)
Providing {\em average\/} link travel times results in the fact that
information on individual fluctuations has been lost.  Using the 30\%
noisy optimal routing algorithm can thus be considered as a heuristic
way to compensate for that.  In general, it is seems from this and
other~\cite{Arthur:bar,Hogg:diversity,Nagel:Rasmussen}
computational experiments that such fluctuations are necessary to
obtain a {\em globally\/} robust outcome.

Note that making the microsimulation feedback more individual and thus
more realistic is easily possible~\cite{Nagel:NRW} and also seems to
make the relaxation more robust.  Changing the TRANSIMS framework
towards such an approach and investigation of the consequences is the
subject of further study.

\section{Summary and Conclusion}

This paper presents a day-to-day re-routing relaxation approach for
traffic simulations.  Starting from an initial planset for the routes,
the route-based microsimulation is executed.  The result of the
microsimulation is fed into a re-router, which re-routes a certain
percentage of all trips.  This procedure is repeated until a certain
amount of convergence is reached.  In this paper, the convergence
criteria was the vanishing of deadlocks in the microsimulation.  It is
shown that this approach makes the traffic patterns in the
microsimulation much more reasonable, in the sense that it gets rid of
heavy congestion in residential areas which are clearly unrealistic.
Quantitative comparisons are in preparation but go beyond the scope of
this paper.  Further, it is shown that the relaxation method described
in this paper can lead to strong oscillations in the solutions.  An
economics/behavioral interpretation of the method may be useful to
find more realistic (and hopefully also computationally more robust)
approaches in the future.

\section*{Acknowledgements}

I thank Paula Stretz for her continuous work and commitment in improving
the TRANSIMS microsimulation.  Deborah Kubicek wrote the graphics package
used for displaying the results.  Many more people on the TRANSIMS team
helped with less visible tasks.

I also thank Maya Paczuski for pointing out that averaging the link
travel times actually {\em destroys\/} information, i.e.\ the
microsimulation feeds back global but still incomplete information.
Steven Durlauf commented on the economics interpretations.  Martin
Pieck thoroughly read through and commented on a draft.

%\bibliographystyle{unsrt}
%\bibliography{kai,ref}

\begin{table}
\tightenlines
\halign{&\hfil#\hfil\cr
\noalign{\smallskip\hrule\smallskip\hrule\smallskip}
iteration\  & \ re-planning percentage\  & \ \# of trips through study
area\  & \ \# of lost vehs. \cr
\noalign{\smallskip\hrule\smallskip}
0 & 20 & 285393 & 44861 \cr
1 & 10 & 275545 & 24709 \cr
2 & 10 & 272199 & 23596 \cr
3 & 10 & 267538 & 36167 \cr
4 & 10 & 265701 & 17969 \cr
5 & 10 & 263255 & 19287 \cr
6 &  5 & 262301 & 15126 \cr
7 &  5 & 261284 & 21867 \cr
8 &  5 & 260155 & 14763 \cr
9 &  5 & 259335 & 16498 \cr
10 & 5 & 258501 & 15025 \cr
\noalign{\smallskip\hrule\smallskip}
}
\vskip1cm
\caption{\label{written}%
Number of trips going through the study area and lost vehicle counts
for different iterations.
}
\end{table}

\begin{figure}
% 1
\caption{\label{procedure}%
Diagramatic view of the relaxation procedure.
}
\end{figure}

\begin{figure}
% 1
\caption{\label{network}%
The so-called ``focussed'' network of the Dallas-Fort Worth area which
is the basis for the study.  Dallas downtown is visible in the
south-east; Fort Worth downtown is recognizable in the south-west from
the shape of the major routes.  Note the rectangular shape
north-north-east of Dallas downtown.  Here, in the study area, the
network consists of {\em all\/} streets, including small residential
ones.  Further out, the digitized road network gets thinner and
thinner.  --- The planner always runs on the whole network visible in
this plot, whereas the microsimulation only runs on the streets of the
study area.
}
\end{figure}

\begin{figure}
% 2
\caption{\label{0-iteration}%
Snapshot of the study area at 10:00am using the initial
planset.  The east-west freeway is the LBJ freeway; the north-south
freeway is the Dallas North Tollway.  
}
\end{figure}

\begin{figure}
% 3
\caption{\label{deadlocks}%
The left figure shows a situation where a complete jam has formed
around a block, and all vehicles at the intersections want to make
right turns, but they are blocked by the last vehicle of the jam in
front.  In the right figure, the black vehicles cannot make their
desired left turn because of vehicles in the desired lanes.  The gray
vehicles cannot make their desired left turns because the current
microsimulation logic demands a gap of at least $v_{max}$ on all
interfering lanes (i.e.\ lanes with priority), yet the black vehicles
are closer than that.  The problem in both figures could be resolved
by making the black vehicles less ``stubborn'', i.e.\ eventually they
decide to go straight or left.  The problem in the right figure could
also be resolved by making the gray vehicles accepting a zero gap in
the interfering lanes when traffic in those lanes is stopped.
}
\end{figure}

\begin{figure}
% 4
\caption{\label{1-iteration}%
Snapshot of the study area at 10:00am after the first iteration.
}
\end{figure}

\begin{figure}
% 5
\caption{\label{10-iteration}%
Snapshot of the study area at 10:00am after the tenth iteration.  Note
that there are no more jams of density one except for short queues at
traffic lights.
}
\end{figure}

\begin{figure}
% 6
\caption{\label{8am}%
Snapshot of the study area at 8:00am after the tenth iteration. 
}
\end{figure}

\begin{figure}
% 7
\caption{\label{9am}%
Snapshot of the study area at 9:00am after the tenth iteration. 
}
\end{figure}

\begin{figure}
% 8
\caption{\label{oscillation4}%
Difference between simulation output based on the 4th iteration
planset and simulation output based on the 3rd iteration planset.
White are locations where the density decreased from the 3rd to the
4th iteration, black are locations where the density increased from
the 3rd to the 4th iteration.
}
\end{figure}

\begin{figure}
\caption{\label{oscillation5}%
Same as Fig.~\protect\ref{oscillation4}, except that it is one
iteration further.  Note that both figures together indicate regular
oscillations.  For example, north of the large intersection in the
middle, density increased from the 3rd to the 4th iteration and
decreased again from teh 4th to the 5th.  Similarly west of that large
intersection and at several other locations.  We used information such
as this to heuristically decrease the re-planning fraction.
}
\end{figure}

\end{document}